\documentstyle[aps,epsf,psfig,rotate,amssymb,multicol]{revtex}
\begin{document}

\title{Irreversible Growth of a Binary Mixture Confined 
in a Thin Film Geometry\\ with Competing Walls}
\author{Juli\'an Candia and Ezequiel V. Albano }
\address{Instituto de Investigaciones Fisicoqu\'{\i}micas Te\'oricas y 
Aplicadas (INIFTA), UNLP, CONICET, \\ 
Casilla de Correo 16, Sucursal 4, (1900) La Plata, Argentina} 

\maketitle

\begin{abstract}
The irreversible growth of a binary mixture
under far-from-equilibrium conditions is studied in
three-dimensional confined geometries of size 
$L_{x} \times L_{y} \times L_{z}$, 
where $L_{z} \gg L_{x} = L_{y}$ is the growing direction. 
A competing situation where two opposite  
surfaces prefer different species of the mixture 
is analyzed. Due to this antisymmetric condition an interface
between the different species develops along the
growing direction. Such interface undergoes a localization-delocalization
transition that is the precursor of a wetting transition in the 
thermodynamic limit. Furthermore, the growing interface also
undergoes a concave-convex transition in the growth mode.
So, the system exhibits a multicritical wetting point where  
the growing interface is almost flat and the 
interface between species is essentially localized at the center
of the film.     
\end{abstract}


\begin{multicols}{2}

The study of interfacial phenomena of materials confined
in thin film geometries under equilibrium conditions 
has drawn enormous attention over the last decades 
\cite{fish,diet,parr}. Interfacial phase transitions 
and critical phenomena in confined samples exhibit a quite
distinct physical behavior compared to that occurring in 
the bulk due to the finite separation between the walls
and the specific wall-particle interaction \cite{fish}.
The understanding of this new type of phenomena is not
only of general and fundamental interest but may also 
play an important role in the development of new technologies.
Accordingly, the properties of films of magnetic materials
\cite{urba,diaz,bind1,alba1}, polymer blends \cite{kerl,muel}, 
binary mixtures \cite{bind2}, fluids \cite{sull}, etc, 
have been investigated thoroughly.
Liquid-gas condensation under confinement between identical walls
displays capillary condensation when the walls promote the fluid 
phase \cite{evan}. Mapping the fluid into
a lattice gas model and using the magnetic terminology, the wall-
particle interaction is accounted for by a surface magnetic field
\cite{urba,diaz,bind1,alba1}. For this reason, 
Ising-like models are very useful to understand the underlying 
physics of more complex systems \cite{footnote}.
Another interesting scenario takes place when an 
Ising film is confined between two competing walls a distance $L$
apart from each other, so that the surface magnetic fields ($H$)
are of the same magnitude but opposite direction.
These competing fields cause the emergence of an interface that
undergoes a localization-delocalization transition at an $L-$dependent
temperature $T_{w}(L,H)$ that is the precursor of the true wetting 
transition temperature $T_{w}(H)$ of the infinite system.

It is surprising that, in contrast to the effort devoted
to the understanding of wetting phenomena under equilibrium conditions,
the non-equilibrium counterpart has received very little attention. 
For instance, Hinrichsen et al. \cite{hinr} have introduced a nonequilibrium 
growth model of a one dimensional interface that undergoes a 
transition from a bounded state to
a non-bounded one. 
 
Within this context, the aim of this work is to perform
an extensive numerical study of the irreversible growth of a magnetic
material confined between parallel walls where competing surface 
magnetic fields act. It is shown that the interplay between confinement
and growth mode leads to a physically rich phase diagram (on the plane 
$H-T$) that exhibits a localization-delocalization 
transition in the interface that runs along the walls and a 
change of the curvature of the
growing interface running perpendicularly to the walls. Extrapolation
of this scenario to the thermodynamic limit leads to a multicritical 
wetting point. 
   
The growth of a ferromagnetic material, with spins having two possible
orientations, is studied using the
so called Magnetic Eden Model (MEM) \cite{mem}.
Monte Carlo simulations are performed
on the square lattice in $(2 + 1)-$dimensions, using a
rectangular geometry $L_{x} \times L_{y} \times L_{z}$ 
with $L_{z} \gg L_{x} = L_{y} = L$.
The location of each site on the lattice is specified 
through its rectangular coordinates $(i,j,k)$,
($1 \leq i,j \leq L$, $1 \leq k \leq L_{z}$).
The starting seed for the growing cluster is a plane 
of $L \times L$ spins placed at $k=1$ and cluster growth takes place
along the positive longitudinal direction (i.e., $k \geq  2$).
The boundary conditions are periodic along one of the transverse 
directions (say in the $i-$direction) but open along the 
remaining transverse direction. In the latter,
competing surface magnetic fields $H_1 > 0$ ($H_L < 0$) acting on
the sites placed at $j=1$ ($j=L$), with $H = H_1 =|H_L|$, are
considered. Then, clusters are grown by selectively adding
spins ($S_{ijk}= \pm 1$) to perimeter sites, which are defined as the
nearest-neighbor (NN) empty sites of the already occupied ones.

Considering a ferromagnetic interaction of strength $J > 0$ 
between NN spins, the energy $E$ of a given configuration of 
spins is given by
$$
E = - \frac{J}{2} \left( \sum_
{\langle ijk,i^{'}j^{'}k^{'} \rangle} S_{ijk}S_{i^{'}j^{'}k^{'}} \right) - 
$$
\begin{equation}
\ \ \ \ \ \ H  \left( \sum_{\langle ik, \Sigma_1 \rangle } S_{i1k} -
\sum_{\langle ik, \Sigma_L \rangle } S_{iLk}  \right)  \ \ , 
\end{equation}
\noindent where the summation $\langle ijk,i^{ '}j^{ '}k^{ '}\rangle$
is taken over occupied NN sites, 
while $\langle ik, \Sigma_1 \rangle$,
$\langle ik, \Sigma_L \rangle$ 
denote summations carried over occupied sites on 
the surfaces $j=1$ and $j=L$, respectively.
Throughout this work we set the Boltzmann constant equal 
to unity
and we take the temperature,
energy, and magnetic fields measured in units of $J$.
The probability for a perimeter site
to be occupied by a spin is taken to be
proportional to the Boltzmann factor 
$\exp(- \frac{\Delta E}{T})$, where $\Delta E$
is the change of energy involved in the addition of 
the given spin. 
At each step, the probabilities of adding up and down
spins to a given site have to be evaluated
for all perimeter sites.
After proper normalization of the
probabilities, the growing site and the orientation of the spin
are determined with Monte Carlo techniques.
Using this procedure, clusters having up to $10^{9}$ spins have  
typically been grown.
Although both the interaction energy and the
Boltzmann probability distribution considered for the MEM are
similar to those used for the Ising model 
with surface magnetic fields [9,12],
it must be stressed that these two models
operate under extremely different conditions, namely the MEM 
describes the {\bf irreversible growth} of a magnetic material 
and the Ising model deals with a magnet under {\bf equilibrium}.
Previous studies have demonstrated that the MEM in $(1 + 1)-$dimensions
is not critical but it exhibits a second-order transition
at $T_{c} = 0.69 \pm 0.01$ in $(2 + 1)-$dimensions \cite{pre}.

The phase diagram of the MEM in a confined geometry with competing
surface fields is very rich and exhibits eight regions. For the
sake of clarity we will first discuss the main
characteristics of each of these regions by means of snapshot
configurations (see figure 1). Subsequently, we will 
quantitatively locate the boundary between these regions and plot the 
corresponding phase diagram (see figure 2). Finally,
we will draw the phase diagram in the thermodynamic limit 
(see inset of figure 2) by extrapolating finite-size results
(see figure 3).  

The vertical straight line at $T_{c}(L=12) = 0.84$ (see figure 2)
represents the $L-$dependent ``critical'' temperature of the
finite system \cite{xx}. So, the left (right) hand side part of
the phase diagram corresponds to the ordered (disordered) growth
regime that involves Regions $I,II,III,IV$, and $A$ 
(Regions $V,VI$, and $B$). 

For low temperatures and small fields 
(Region $I$ of figure 2, see also figure 1(a)), the 
system grows in an ordered state and a domain having mostly spins 
with a single orientation prevails. 
Due to thermal fluctuations, small 
clusters with the opposite orientation may appear,
preferably on the surface where the field competing with the 
dominant orientation is applied. These ``drops'' might grow
and when the typical size of the fluctuation is of the order 
of $L$, a magnetization reversal may occur, 
thus changing the sign of the dominant domain.
The formation of sequences of well ordered domains is a finite
size effect that is also
characteristic of the ordered phase of some confined spin systems
such as the Ising magnet \cite{alba1}. 
Due to the open boundary conditions, 
empty perimeter sites at the confinement walls 
experience a missing neighbor effect. 
Since $H$ is too weak to compensate this effect, 
the system grows preferentially along the center of the sample 
as compared to the walls, and the
resulting growth interface exhibits a convex shape.
So, Region $I$ corresponds to the Ising-like nonwet
state and the convex growth regime.

\begin{figure}
\centerline{
\epsfxsize=8cm
\epsfysize=6.5cm
\epsfbox{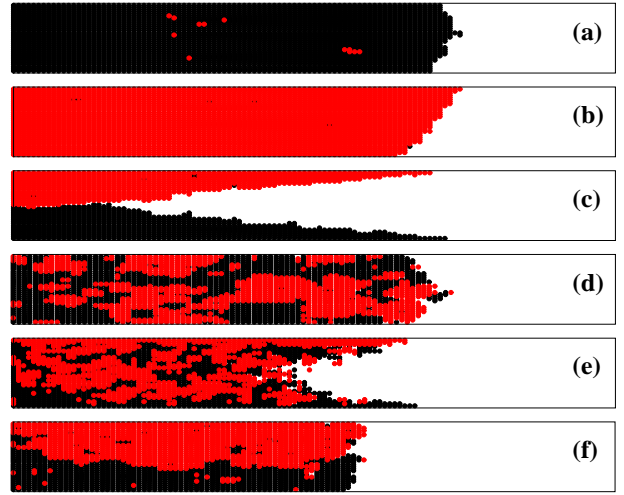}}
\vskip 0.5 true cm 
\caption{Snapshot pictures showing a longitudinal slice 
given by a fixed value of the transverse coordinate $i$. 
Red (black) circles correspond to spins up (down). 
The surface field on the upper (lower) confinement
wall is positive (negative). 
The snapshots correspond to a lattice size $L=32$ and 
several different values of temperature and surface fields: 
(a)$H=0.05$, $T=0.6$; (b)$H=0.5$, $T=0.55$;
(c)$H=1.4$, $T=0.6$; (d)$H=0.1$, $T=1.0$; (e)$H=1.6$, $T=1.4$;
and (f) $H = 0.20$, $T = 0.82$.}
\label{Figure1}
\end{figure} 

Increasing the field but keeping the temperature low enough, 
one may enter Region $II$ (figure 2, see also figure 1(b)). 
Here the system is still in its ordered phase and
neighboring spins grow preferably parallel-oriented. The surface fields
in this region are stronger and thus capable of compensating the missing 
NN sites at the surfaces. But, since the fields on both surfaces have opposite
orientations, one has that, on the one hand, the field that has the 
same orientation as that of the dominant spin cluster will favor 
the growth of surface spins, 
while on the other hand, the sites on the surface with opposite field will 
have a lower growing probability. 
Thus, on this disfavored side the growing interface 
becomes pinned and 
the curvature of the growing interface is not defined. 

Keeping $H$ fixed within Region $II$ but increasing the temperature, 
thermal noise will enable the formation of drops on the disfavored side 
that eventually may nucleate into larger clusters as the temperature 
is increased even further. This process may lead to the emergence of 
an up-down interface, separating oppositely oriented domains, running in the
direction parallel to the walls. Since sites along
the up-down interface are surrounded by oppositely oriented NN spins, 
they have a low growing probability. 
So, in this case the system grows preferably along the confinement 
walls and the growing interface is concave (figure 1(c)). 
Then, as the temperature is
increased, the system crosses to Region $A$ (see figure 2) and we observe 
the onset of two competitive growth regimes: {\it (i)} one  
exhibiting a non-defined growing curvature that appears when 
a dominant spin orientation 
is present, as in the case shown in figure 1(b); 
{\it (ii)} another that appears
when an up-down interface is established and the system
has a concave growth interface, as is shown in figure 1(c). 

\begin{figure}
\centerline{
\epsfxsize=8cm
\epsfysize=6cm
\epsfbox{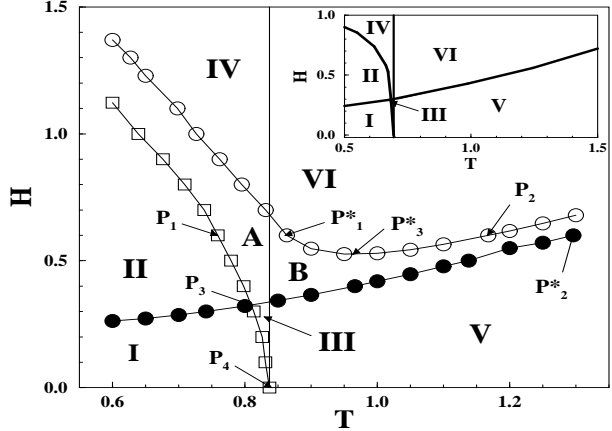}}
\caption{$H-T$ phase diagram corresponding to a lattice of size $L=12$. 
The vertical straight line at $T_{c}(L) = 0.84$ corresponds 
to the $L-$dependent critical temperature [17], which
separates the low-temperature ordered phase
from the high-temperature disordered phase.
Open (filled) circles refer to the transition between non-defined and 
concave (convex) growth regimes, and squares stand for the Ising-like 
localization-delocalization transition curve.
Eight  different regions are distinguished, 
as indicated in the figure. 
Also indicated are seven representative points that are 
discussed in the text.
The inset shows the phase diagram
corresponding to the thermodynamic limit composed of
six different regions.}
\label{Figure2}
\end{figure} 

Further increasing the temperature and for large enough fields, 
the formation of a stable longitudinal up-down interface
that pushes back the growing interface is observed. So,
the system adopts the concave growth regime (see figure 1(c) corresponding to
Region $IV$ in figure 2). 
Increasing the temperature beyond $T_c(L)$,
a transition from a low-temperature ordered 
state (Region $IV$) to a high-temperature disordered state
(Region $VI$, see figure 1(e)), both within the concave 
growth regime, is observed. Analogously, for small enough fields, 
a temperature increase drives the system from the ordered convex growth
regime (Region $I$) to the disordered convex growth regime (Region $V$,
see figure 1(d)). As shown in figure 2, there is also an 
intermediate fluctuating state
(Region $B$) between Regions $V$ and $VI$, 
characterized by the competition between
the disordered convex growth regime and the disordered concave one.  

Finally, a quite unstable and small region (Region $III$) that
exhibits the interplay among the growth regimes of the contiguous
regions, can also be identified. Since the width of Region $III$
is of the order of the rounding observed in $T_{c}(L)$, large
fluctuations between ordered and disordered states are observed,
as well as from growth regimes of non-defined curvature to  
convex ones. However,
figure 1(f) shows a snapshot configuration that is the fingerprint
of Region $III$, that may prevail in the thermodynamic limit,
namely a well defined spin up-down interface with an almost
flat growing interface. 

Let us now locate the transition curves between the 
already discussed regions. For this purpose, the mean transverse
magnetization $m(k,L,T,H)$ is defined as
\begin{equation}
m \ \ (k,L,T,H) = \frac{1}{L^2} \sum_{i,j = 1}^{L} S_{ijk} ,      
\end{equation}
\noindent and the susceptibility ($\chi$)
is defined in terms of the magnetization fluctuations.
Then, using a standard procedure\cite{alba1}, 
the  localization-delocalization ``transition''  
curve corresponding to the up-down interface running along the walls
can be obtained considering that on the $H-T$ plane,
a point with coordinates $(H_w,T_w)$ on this curve
maximizes $\chi(H,T)$.
The obtained curve for
the confined system is shown in figure 2. This
localization-delocalization ``transition'' is also known as 
a pre-wetting ``transition'' and actually becomes a true 
wetting transition in the thermodynamic limit. 

Since the MEM is a kinetic growth model,
another kind of nonequilibrium wetting ``transition'' associated 
with the curvature of the growing interface can also 
be identified. In fact, varying the surface fields one 
finds a transition between a wet state (that corresponds 
to a concave interface) and a nonwet state 
(associated with a convex interface), as already observed in figure 1. 
Clearly, two different contact
angles must be defined in order to locate this transition,
namely $\theta_D$ for the angle corresponding to the 
dominant spin cluster, and $\theta_{ND}$ for the one that 
corresponds to the non-dominant spin cluster.
Both contact angles can straightforwardly be determined by measuring
the location of the growth interface averaged over a
sufficiently long growing time.  
In this way,  three  different  growth  regimes  can  be  distinguished: 
 {\it (i)} the concave growth regime that occurs when the system wets the 
walls on both sides (i.e. for $\theta_D, \theta_{ND} < \frac{\pi}{2}$),
{\it (ii)} the convex growth regime that 
occurs for $\theta_D, \theta_{ND} > \frac{\pi}{2}$, 
and {\it (iii)} the regime of non-defined
curvature that occurs otherwise. The corresponding transition curves
obtained for the confined system are shown in figure 2.

We will now extrapolate our results to
show that the rich variety of phenomena
found in a confined geometry is 
still present in the $L \rightarrow \infty$ limit,
leading to the phase diagram shown in the inset of figure 2.  
In order to illustrate the extrapolation procedure, 
the following seven representative points of the
finite-size phase diagram are discussed in detail: 
{\it (i)} the points labeled $P_1$, $P_1^*$, $P_2$, 
and $P_2^*$, that correspond to the
intersections of the $H = 0.6$ line with the various transition
curves shown in figure 2, and {\it (ii)} the points labeled
$P_3$, $P_3^*$, and $P_4$, that refer to 
the intersection point between Regions 
$I, II, III$, and $A$, the minimum of the limiting curve between 
Regions $IV$-$VI$ and $A$-$B$, and 
the zero-field transition point, respectively.

\begin{figure}
\centerline{
\epsfxsize=7.5cm
\epsfysize=5.0cm
\epsfbox{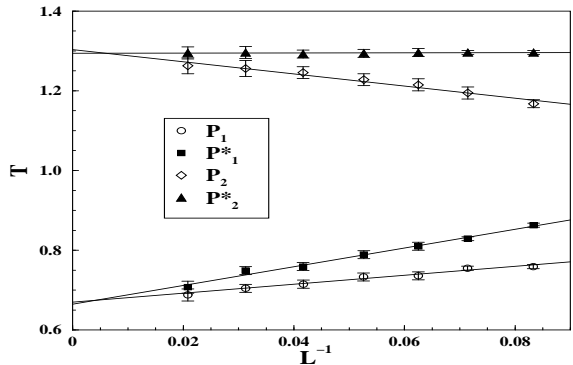}}
\caption{Plots of $T$ versus $L^{-1}$ 
corresponding to the points $P_1,P_1^*,P_2$, and $P_2^*$, 
all of them with $H = 0.6$. 
The fits to the data (solid lines) show 
that $P_i \rightarrow P_i^*$  $(i=1,2)$ for
$L \rightarrow \infty$.}
\label{Figure3}
\end{figure} 

Figure 3 shows plots of $T$ versus $L^{-1}$  
corresponding to the points $P_1,P_1^*,P_2,$ and $P_2^*$. 
The results from the extrapolations are: 
$T_1=0.67\pm0.01$, $T_1^*=0.66\pm0.01,$
and
$ T_2=1.30\pm0.02$, $T_2^*=1.29\pm0.01,$
pointing out that, within error bars, 
$P_i \rightarrow P_i^*$  $(i=1,2)$ in the $L \rightarrow \infty$
limit. Using the same procedure,
the extrapolations of $P_3$ and $P_3^*$
(not shown here) give:
$H_3=0.30\pm0.01$, $H_3^*=0.31\pm0.02,$
and $T_3=0.69\pm0.01$, $T_3^*=0.71\pm0.03$.
So, one has $P_3 \rightarrow P_3^*$ for $L \rightarrow \infty$  
within error bars. Finally, the extrapolation of $P_4$ is
$T_4 =  T_c = 0.69\pm0.01$.
  
Using the above-mentioned extrapolation procedure,
the phase diagram in the thermodynamic limit can be drawn,  
as shown in the inset of figure 2. 
By comparison with the finite-size phase diagram of
figure 2, notice that, besides the fact that
the crossover Regions $A$ and $B$ vanish in this limit,
it is expected that Region $III$ may remain. 
Although this (very tiny!) region corresponds to
a physically well characterized growth regime, since
one expects that the system in this region may grow in an ordered phase
with a localized up-down domain interface and a convex growing interface,
statistical errors due to large fluctuations close to criticality 
hinder the exact location of this region. 
 
Comparing the equilibrium wetting phase diagram of the Ising model
\cite{parr,bind1,alba1} and that
of the MEM, it follows that the non-equilibrium
nature of the latter introduces new and rich  physical features of interest:
the nonwet (wet) Ising phase splits out into Regions $I$ and $II$
(Regions $III$ and $IV$), both within the ordered regime ($T < T_{c}$) 
but showing an additional transition in the interface growth mode. 
Also, the disordered state of the Ising system ($T > T_{c}$) splits
out into Regions $V$ and $VI$ exhibiting a transition in the 
interface growth mode. The interplay between wetting of two
interfaces, one of them defined between differently
orientated domains that runs along the film, and the remaining one 
resulting from the growing process, leads to a multicritical
wetting point close to $H^{MC} = 0.31 \pm 0.01, T^{MC} = 0.70 \pm 0.02$. 
At this particular multicritical point both interfaces are essentially
perpendicular to each other (figure 1(f)).

We hope that the presented results will, on the one hand, contribute to 
the understanding of 
the rich and complex physical phenomena exhibited by the
irreversible growth of binary mixtures in confined
geometries, and on the other hand, stimulate 
further experimental and theoretical work. 
       
{\bf Acknowledgments}: This work was supported by CONICET, 
UNLP, and ANPCyT (Argentina).

\end{multicols}

\begin{references}
\bibitem{footnote} From now on we will unify the discussion in terms of
a magnetic language. However, the relevant physical concepts can be extended 
to other systems such as fluids, polymers, and binary mixtures.
\bibitem{xx} Among various definitions, $T_{c}(L)$ 
has been here identified with the peak of the susceptibility.
\end{references}
\end{document}